\documentclass[conference]{IEEEtran}
\usepackage{blindtext, graphicx}
\usepackage{lettrine}
\usepackage{fancyhdr} 
\usepackage{cite}
\usepackage{float}
\let\labelindent\relax
\usepackage{enumitem}

\ifCLASSINFOpdf
\else
\fi
\begin{document}
%
\title{Automating Network Error Detection using Long-Short Term Memory Networks}


\author{\IEEEauthorblockN{Moin Nadeem\IEEEauthorrefmark{1},
Vibhor Nigam\IEEEauthorrefmark{2},
Dimosthenis Anagnostopoulos\IEEEauthorrefmark{3}, and
Patrick Carretas\IEEEauthorrefmark{4}}
\IEEEauthorblockA{Smart Network Platform\\
Technology, Product, \& Experience \\
Philadelphia, Pennsylvania\\ \\ Contact: moin\_nadeem@comcast.com\IEEEauthorrefmark{1}, vibhor\_nigam@comcast.com\IEEEauthorrefmark{2},\\ dimos\_anagnostopoulos@comcast.com\IEEEauthorrefmark{3}, patrick\_carretas@comcast.com\IEEEauthorrefmark{4}}


}

%


\maketitle

\begin{abstract}
In this work, we investigate the current flaws with identifying network-related errors, and examine how much potential K-Means and Long-Short Term Memory Networks may have in solving these problems. We demonstrate that K-Means is able to classify messages, but not necessary provide meaningful clusters. However, Long-Short Term Memory Networks are able to meet our goals of providing an intelligent clustering of messages by grouping messages that are temporally related. Additionally, Long-Short Term Memory Networks can provide the ability to understand and visualize temporal causality, which may unlock the ability to warn about errors before they happen. We explore the impact of this research, and provide some suggestions on future work.
\end{abstract}


%
\IEEEpeerreviewmaketitle

\section{Introduction}  

\lettrine{T}{he} performance of any network provider, and ultimately, its revenue stream, is tied to the performance of its products. If one is able to lower the mean time towards identifying, classifying, and resolving network errors, they are able to provide immense value by potentially mitigating penalties from \textit{Service Level Agreement (SLA)} violations, thus increasing customer retention, improving their reputation, and cutting down on increased human resources costs.

Recognizing these benefits, the networking community has proposed a number of architectures, frameworks, and protocols to increase network reliability. Recent efforts on improving network reliability have converged around examining network topology, improving the random coding of the content from a packet level, or applications of percolation theory \cite{Yang2016}. However, system logs left on network devices, which have previously proved invaluable \cite{Turner2013} \cite{Turner2010}, have been largely ignored by the greater community. 

Intuitively, this may be explained by the fact that router system log (syslog) messages are primarily designed for router vendors to track and debug their own software and hardware problems. This is in contrast to network service providers utilizing them for monitoring network health and troubleshooting performance anomalies. As a result of the fact that these messages are designed for the vendor and not the customer, these messages pose some challenges from a data consumption and data consistency standpoint. 

To begin, they are essentially free-form text with little to no shared structure. The breadth of devices available makes it difficult to compare and contrast information that is logged among various vendor and router operating system formats. 

Secondly, as these devices are embedded hardware devices, the system logs often contain information that is too low-level for most purposes. This makes it difficult to directly translate these logs into network events without some domain-level expertise to meaningfully abstract and aggregate this data. 

Lastly, the information-to-noise ratio of syslog messages is quite low; most syslog messages are generated purely for debugging purposes, and have little-to-no relation to the status of the network. This makes it difficult to intercept information from a heavy stream of millions of messages per second, and determining dependencies between messages at this information-to-noise ratio has proven to be quite challenging.

Therefore, although large ISPs (Internet Service Providers) networks consist of thousands of routers delivering millions of information-rich syslog messages per day, the lack of uniform structure and the ambiguity in relational structure makes it difficult to utilize system log messages in network performance applications. Instead, most ISPs require the input of a network engineer with domain expertise to focus on a rather small, yet important, subsets of messages. Regular Expressions (RegEx) \cite{Thompson1968} are commonly used to capture pre-determined patterns from syslog messages, and this output is used to trigger alarms and issue operational tickets for engineers to diagnose. 

However, this problem in itself has a few difficulties. The parsers and relational models with the messages needs to be constantly updated to reflect new errors, new router models, and new operating systems. Additionally, network issues which have not been already captured by a regular expression may fly under the radar. Furthermore, network engineers have to focus on a particular router and a narrow time-window when examining the raw system log messages, which creates a very time consuming and arduous process if the particular incident spans hundreds of routers and thousands of messages. This makes it easy to lose sight of the "big picture" as far as the frequency and scope of the investigation requires when debugging specific network problems. Therefore, it becomes clear that the ability to system log messages in a more intelligent fashion has a strong impact, particularly if one is able to automatically able to cluster related errors.
\subsection{Aims}
In this work, we focus on proactively mining information from system log messages, and transforming it into actionable information for software engineers. We design an automated system that transforms and compresses large amounts of low-level, minimally structured data and groups them into a few small high-level network events. This approach is vendor and network independent, does not require domain expertise to update rules, and operates based off of actual network behaviors rather than expected behaviors. In addition, given any particular system log message $x$, we are able to determine which messages were most likely to cause $x$, and which messages may be caused by $x$. In other words, we're able to provide a high-level understanding of how these messages are related temporally.

This permits network operators to quickly identify and focus on important network events. More importantly, this also enables complex network troubleshooting, in which we have the ability to collapse thousands to tens of thousands of system log messages on multiple routers, and we provide the foundations potentially relate these messages to a single network event. This not only creates better infrastructure as far as the ability to monitor overall network health, but also permits increased network visualization. 

As a result, we believe this increased network visualization will lead to a superior understanding of the big picture than raw system logs are able to provide. Lastly, it is worth noting that our approach has the ability to learn errors from the data. This permits us to spot a wider variety of errors, since we're providing insights on "how the network behaves", rather than some preconceived notion of "how the network should behave".

{\centering To summarize,

}

In this work, we take the first steps towards building a general intelligent log analysis framework that has two goals: 
\begin{enumerate}
    \item Insightfully group messages into more meaningful network events, such that system log messages that are related to each other will appear within the same grouping.
    \item Understand the relation between network system log messages, such that we may understand which messages are most likely preceded any message, and which messages are most likely to be caused by it.
\end{enumerate}

The remainder of this paper is organized as follows. Section II provides a background on the relevant technologies at hand. Section III describes the methodologies we used to setup our experiment, and Section IV describes the results we discovered with respect to our operational goals and our dataset. We present some potential next steps for this work in Section V. Finally, Section VI concludes this paper.

\section{Method}
In this section, we discuss our methodologies, and the intuition we had behind them. 

\subsection{KMeans \& HDBScan}
\subsubsection{Preprocessing}
In this study, we gathered information from our Splunk database. This database was developed by gathering information from an amalgamation of system log routers, and included information such as:
\begin{itemize}
    \item \textit{sender\_message}, the syslog message from each router.
    \item \textit{sender\_ip}, the IP address for each router. This field may serve to group routers together if necessary.
    \item \textit{sender\_facility\_num}, an integer representing the facility that these routers were in. This field may serve to group routers together by physical location if desired.
    \item \textit{sender\_level\_num}, a priority level for each message sent.
\end{itemize} 

For our purposes, we only fed the KMeans \& HDBScan algorithm messages with a \textbf{priority less than 5}. An explanation of the priority levels is given below, and will serve to be important for the Long-Short Term Memory Networks.
\begin{enumerate}
    \item Emergency: the system is unusable.
    \item Alert: action must be taken immediately.
    \item Critical: critical conditions.
    \item Error: error conditions.
    \item Warning: warning conditions.
\end{enumerate}

We generated a subset of these messages from Splunk which arrived within fifteen minutes of each other. This resulted in 2,330,225 system log messages. Next, we performed the following preprocessing steps. 

We removed any row from the dataset which presented incomplete information. We defined incomplete information as the lack of any field from the ones enumerated above. This removed 438 rows.

Next, we ensured that the distribution of timestamps was uniform throughout the process. Since we are dealing with a timeseries dataset, having an even distribution of data helps us ensure that our system log messages are related to one another, and therefore may be clustered properly. 

Upon examination of the \textit{sender\_unixtime} property, we noticed several timestamps from March 2017, when the bulk of our dataset was from January 2018; a cause for concern due to the fact that we had selected only dates from Monday, January $8^{th}$ on Splunk. After speaking with Network Engineers, we determined that the clock on the router is misconfigured, and dropped the corresponding rows. 

Next, we bucketed the messages into five minute intervals with the intuition that system log messages more than five minutes apart from each other have very little relation between them. This yielded 202 buckets, and we chose one with 51,040 messages to explore. A distribution of buckets is visible in Figure 6.

\begin{figure}
    \centering
    \includegraphics[width=0.5\textwidth]{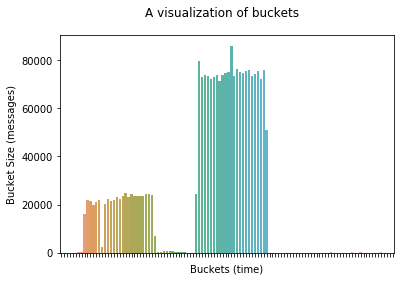}
    \caption{A visualization of each bucket as a function of time. You can see there are two clusters of time, and we have selected a bucket that is halfway between them.}
    \label{fig:my_label}
\end{figure}

However, these clustering algorithms are fed points in $n$-dimensional space, not words. Therefore, we vectorized our words by creating one dimension for every unique word from all of the messages, and set the value of the $n^{th}$ dimension equal to how many times some particular word occured in that message. This left us with a dataset that was a matrix of shape (80000, 2243).
\subsubsection{Algorithm} 
In both approaches, we apply a standard approach to each algorithm. For K-Means, we use the \textit{sklearn} package, and we use the \textit{hdbscan} package as available on PyPi. In order to determine the $k$ in k-means, we applied the Bayesian Information Criterion \cite{Neath2012}, which permits us to score any particular model. This, in turn, permits us to determine which $k$ may provide the lowest score. The impact of this will be discussed in Section IV.

\subsection{LSTM}
\subsubsection{Preprocessing}
For this section of the study, Dimosthenis Anagnostopoulos leveraged the company's syslog broker to gather a dataset. The data included the following:
\begin{itemize}
    \item \textit{IP}, the IP address of the router at hand.
    \item \textit{Time}, a ISO-formatted timestamp of the message, as sent by the router.
    \item \textit{message\_words}, the syslog message
    \item \textit{Priority}, an integer-based Priority scale as detailed in the \textit{KMeans} section of this paper.
    \item \textit{Hostname}, a FQDN (fully qualified domain name) specifying the physical location of the router.
    \item \textit{ifdescr}, a string containing information about which interface sent the message.
    \item \textit{vendor}, the vendor of the respective router
    \item \textit{sp\_device\_type}, a low level location metric.
    \item \textit{market\_site}, a string containing information about the physical location of our router. This is used to group routers that may contain similar errors.
\end{itemize}

Once we had a copy of the dataset, we removed integers, special characters, both IPv4 and IPv6 addresses, dates, times, usernames, and any additional information in the message after fifty characters. At this point, our dataset contained 347,264 messages.

As we did previously, we removed all outlying timestamps until we had a sufficiently uniform distribution. In this case, 2,000 messages were removed, leaving us with 345,249 messages occurring on January 12th, 2017 from 8:20PM to 8:32 PM. The distribution can be viewed in Figure 7, note how one particular timestamp doesn't dominate the distribution.

Among the 347,264, there existed 1,288 distinct messages among the set, of which only 5 occurred more than 1\% of the time. We ran experiments on both the 1,288 and the 5 messages, which will be detailed later during the results section. 
\subsubsection{Model architecture}
In traditional Machine Learning, the effort is mostly spent on feature extraction. However, in Deep Learning (which uses Neural Networks), most of any individual's effort will be spent fine-tuning their model. As the scope of our investigation was limited to a Proof of Concept rather than creating a model that will be used in production, we focused on determining what underlying concepts of neural networks applied to our research, and which concepts other employees could take forward.

We transformed our data into a many-to-one model in which many previous timesteps are used to predict the next timestep. For our application, this translates to the using the previous $n$ messages to predict the next message. $n$ is a parameter to be tuned, which we will further elaborate on in Section IV.

\begin{figure}
    \centering
    \includegraphics[width=0.5\textwidth]{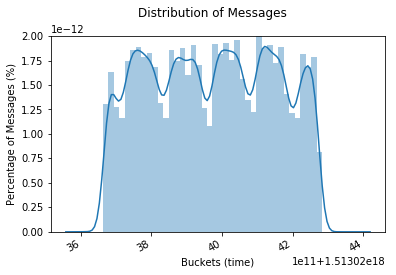}
    \caption{A distribution of timestamps for the LSTM timeseries data.}
    \label{fig:my_label}
\end{figure}

In the last step of our preprocessing, we one-hot encoded our messages. For the uninitiated, one-hot encoded vectors are used to encode categorical variables into a list of numbers, where any particular index is $on$, or $1$ if that category is represented in any particular timestemp. For example, a one-hot encoding of \textit{blue}, if the categories were \{\textit{red, blue, green}\}, would be [0, 1, 0]. This helped create our final dataset, which was of shape (samples, sequence\_length, classes). For our input, such as a sequence of 15 on 1,288 distinct messages, our input was (347264, 15, 1228). We used both Keras' Sequential and Functional API to create a Deep Neural Network, which we then used to train our weights and to fit our output. 

Lastly, in order to deal with high levels of class imbalance, we commonly used \textit{class\_weights}, a parameter in Keras which penalizes certain examples mistakes $x$ times more than others. For example, if we have two classes, \{$a$, $b$\}, where $a$ has 5000 examples, and $b$ has 2500 examples, we may provide a class weight of \{1.0, 2.0\}. This has the effect of making mistakes on $b$ twice as "painful", which helps it learn a better underlying representation.

\begin{figure}
    \centering
    \includegraphics[height=0.40\textwidth]{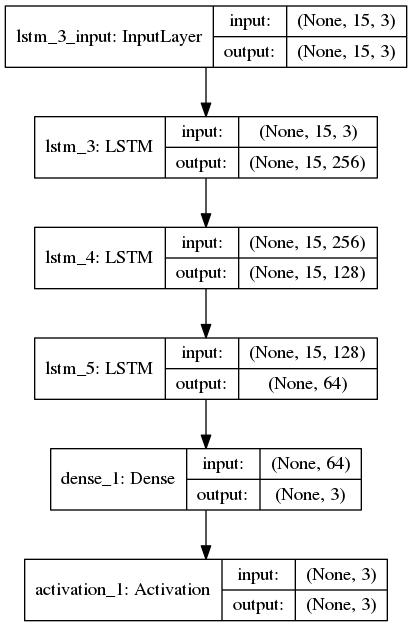}
    \caption{A visualization of our model architecture.}
    \label{fig:my_label}
\end{figure}

\section{Results}
In this section, we investigate and discuss the accuracy and precision of all three of our algorithms, and understand the results of the hidden states for implementation of LSTMs. We shall define successful results as the following:
\begin{itemize}
    \item Creating a system that is able to group errors intelligently. We define intelligence as the ability to understand long-term causal dependence; we identify these as the grouping of patterns from network routers during different time intervals. 
    \item Operates quickly: our algorithms should be able to cluster about fifteen minutes of data from all routers in under a minute. In other words, we would like our algorithm to run in no more than $O(n^2)$.
    \item Permits us to potentially understand how errors are related between one another. In other words, given any particular error, how can we determine what led to it? 
\end{itemize}
\subsection{KMeans}
\begin{figure}
    \centering
    \includegraphics[width=0.45\textwidth]{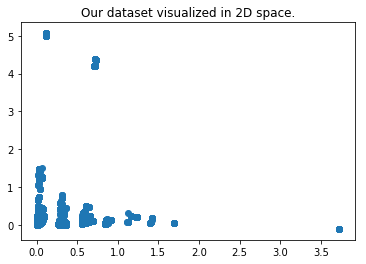}
    \caption{A visualization of the number of message space squeezed to two dimensions. Here, it becomes clear that there do exist distinct clusters.}
    \label{fig:my_label}
\end{figure}
We chose KMeans as one of the first algorithms we worked with due to previous experience, its notoriety as being a versatile algorithm, and its speed. Since this is not a supervised problem (ie. we don't have predetermined labels for each message), we aren't able to provide quantitative metrics. Our examination will be solely qualitative with respect to our goals.

After a discussion with network engineers, we had determined that this dataset will contain no more than 30 clusters due to the natural grouping of errors that was present. This number was also confirmed through a process of trial and error on the number of clusters that would work well. Figure 9 shows a reduced two-dimensional representation of the message space, from which we can discern at least 13 unique clusters \textit{in two-dimensional space}.

There does exist a method to determine an optimal number of cluster called the Bayesian Information Criteron \cite{Neath2012}. Essentially, the Bayesian Information Criterion is a method to score any given model. It does so by attempting to maximize the likelihood that the data follows a normal distribution, while penalizing the score with respect to model complexity. Therefore, it attempts to find the best model that does not overfit.

However, when we implemented the BIC, it tended towards models that gave every message its own cluster. This makes intuitive sense: the BIC attempts to model data into a normal distribution, as well as provide the least complex model possible. In this case, it decided that the least possible complex model was one where each message had its own cluster, indicating that traditional methods for determining the number of clusters are unsuccessful. Shown in the next two pages are tables which serve to provide examples of messages that were clustered together. It is important to observe the common \textit{words in the same cluster}, as it is attempting to cluster similar messages together which may represent some higher-level error.
\onecolumn
\begin{figure}[]
    \centering
    \section{\large Results: K-Means Clustering}
    \caption{}
    \includegraphics[width=\textwidth]{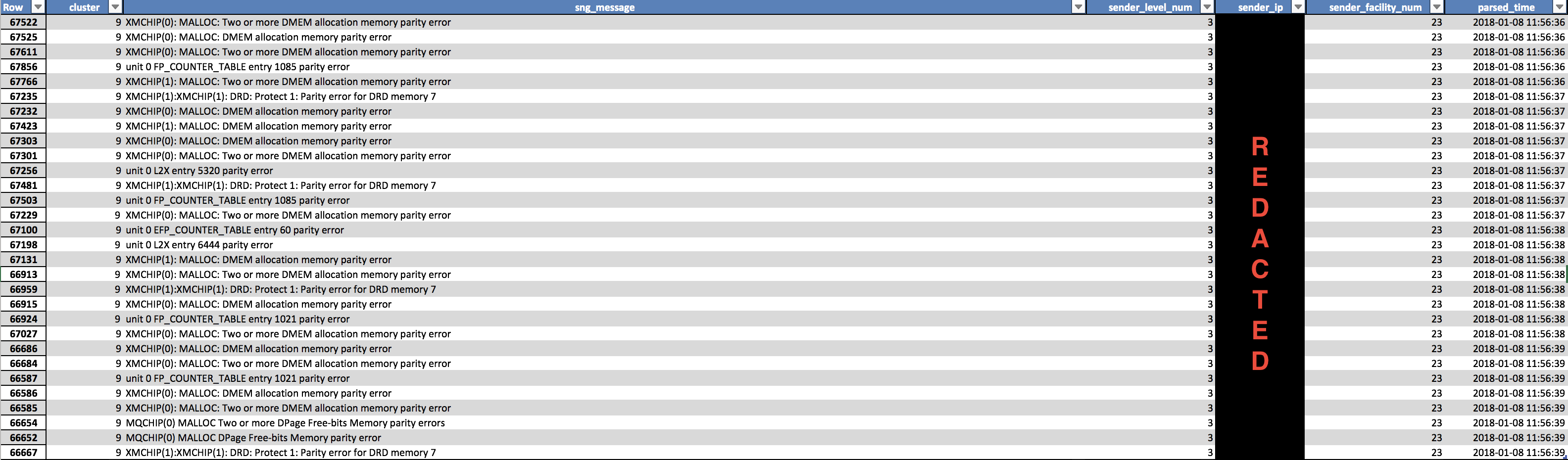}
    
    \centering
    \includegraphics[width=\textwidth]{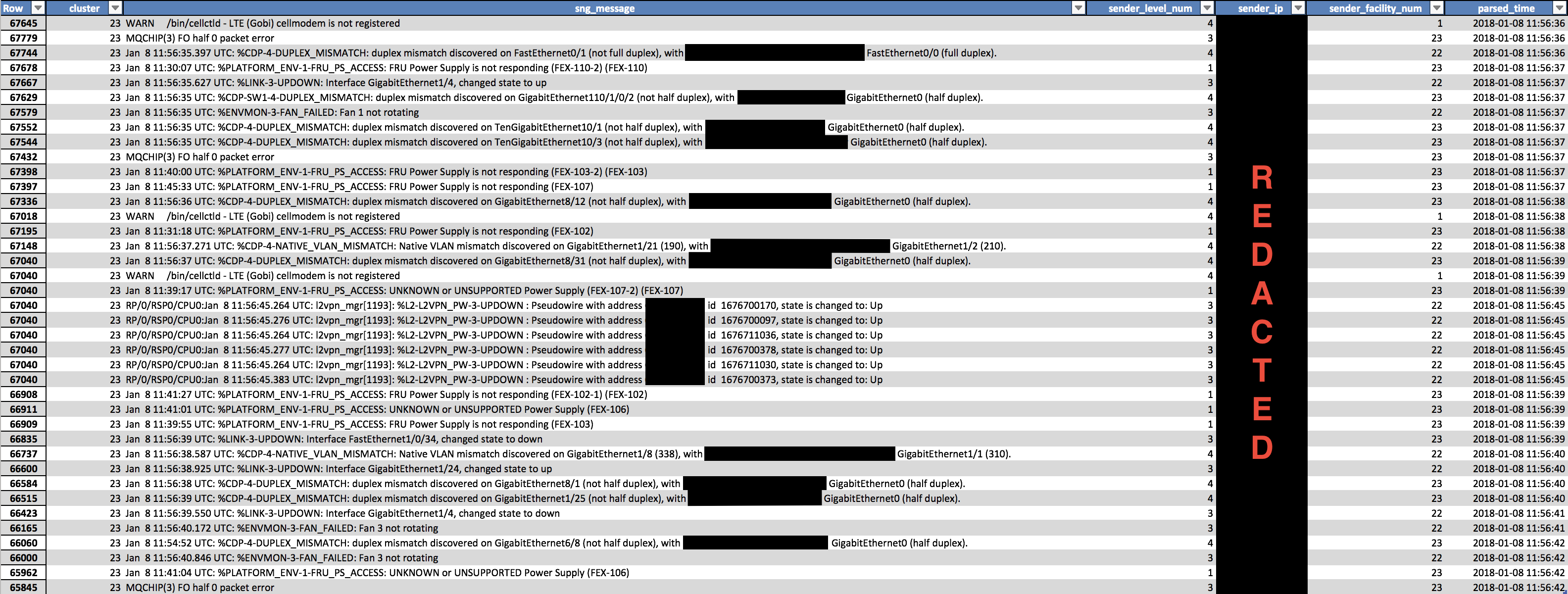}
    
    \centering
    \includegraphics[width=\textwidth]{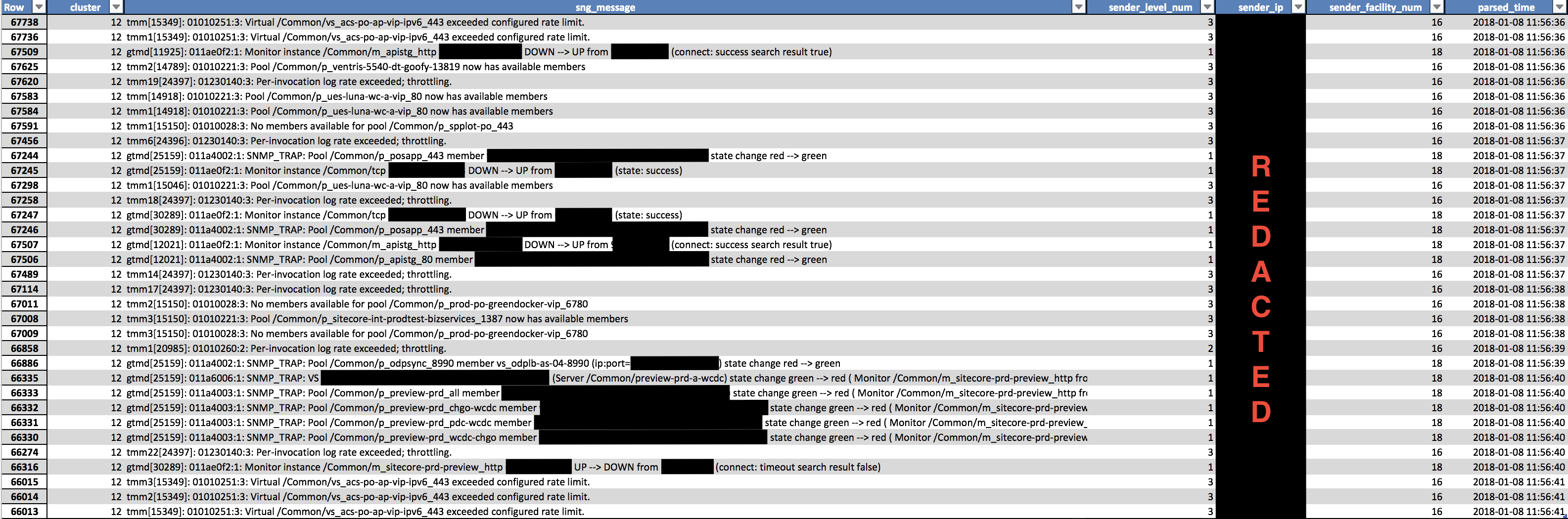}
    
    \centering
    \includegraphics[width=\textwidth]{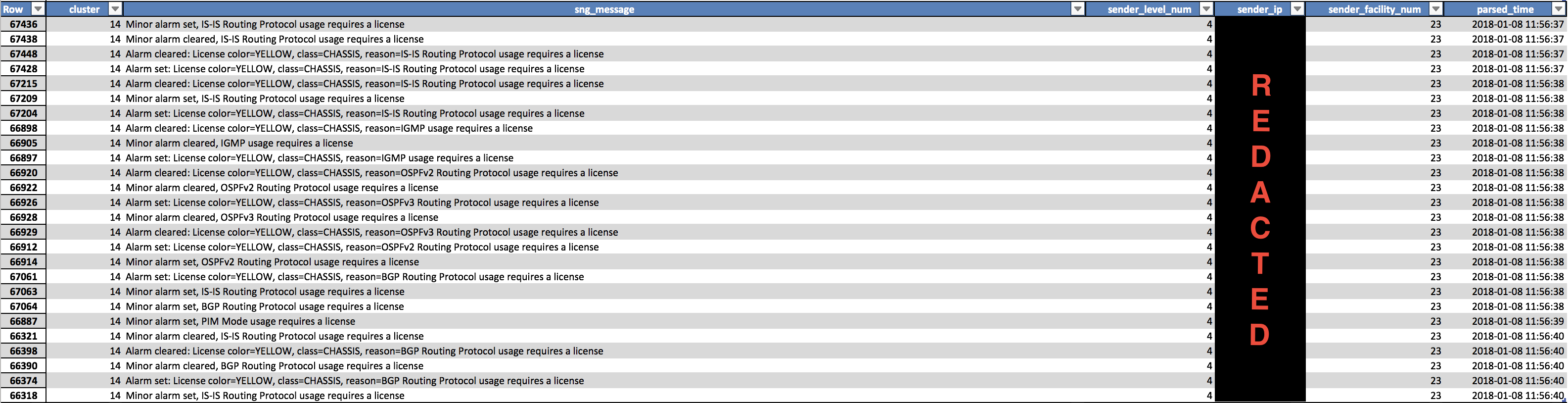}
\end{figure}
\begin{figure}
    \centering
    \section{\large Results (Part 2): K-Means clustering}
    \caption{}
    \includegraphics[width=\textwidth]{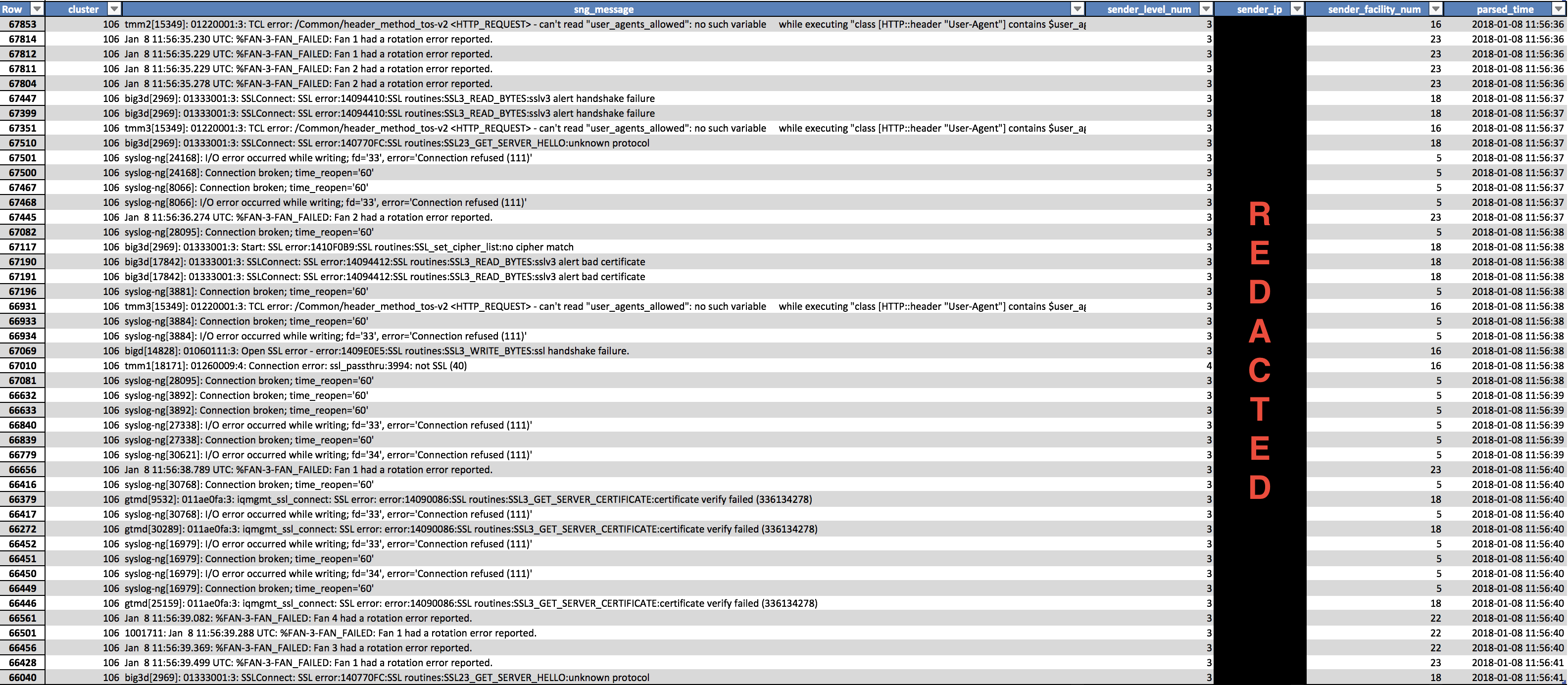}
\end{figure}

\twocolumn

\subsection{KMeans (cont.)}
\begin{figure}
    \centering
    \includegraphics[width=0.4\textwidth]{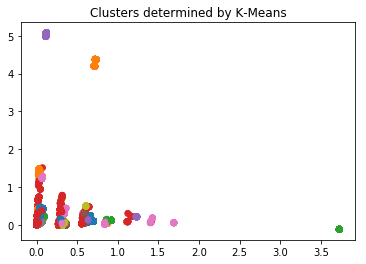}
    \caption{A visualization of clustering as determined by K-Means. Each color represents a different cluster.}
    \label{fig:my_label}
\end{figure}

Returning to our criteria of success for our models, we had to find a model that was efficient, and created some \textit{intelligent} grouping of errors. KMeans certainly achieves the first mark in terms of speed - it is able to cluster up to 1 million data points in 30 seconds. However, this should not be an area of concern due to the fact that at least 5 minutes worth of data will be run at any given time.

However, it is important to note that the clusters themselves are not particularly useful - out of 30 clusters, only six showed slight promise, which are listed above. Let's go ahead and gain a better understanding of each cluster.
\begin{itemize}
    \item Cluster 9 grouped together various memory errors. This is a good start, but is only one example out of a set of many. Let's see what we have from the other clusters.
    \item Cluster 23 has grouped together \textbf{both} power supply errors, network state messages, fan errors, and mismatches together. After discussing this with a network engineer, we determined there was no correlation between these errors. 
    \item Cluster 12 grouped together messages relating to pooling nodes together.
    \item Cluster 14 grouped together alarm messages. As alarms may be detected with simple regex, this is of little use to us. This has the same implications as cluster 9; that is, this tool proves better if we wanted to group message types together.
    \item Cluster 106 grouped together connection level errors, along with fan errors together. This is indicative of the fact that KMeans is not understanding these errors correctly, and will be touched upon later.
\end{itemize}

After examining these errors, it is clear that KMeans is not a valid algorithm to achieve our objective of grouping together errors at a more abstract level. We will next attempt to understand why. Going back to Section II.A.1, we understand that KMeans has the following properties, and its implications on our work: (1), it assumes that all clusters will occur with the same probability. For us, this means that higher-level errors must occur with the same probability, which is inherently untrue. (2) clusters should be mutually exclusive; the ability for a point to be in one cluster means that it must not be in another. (3), we must have relatively clean data; many outliers will confuse the algorithm. 
\begin{figure}
    \centering
    \includegraphics[width=0.4\textwidth]{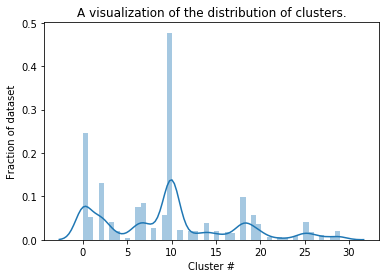}
    \caption{A visualization of the number of clusters, and the probability of each cluster occurring.}
    \label{fig:my_label}
\end{figure}
However, our data inherently violates the first assumption, and just because the fact that a point is in one cluster doesn't mean it shouldn't be in another. In other words, we should be able to associate some syslog messages with different high-level errors, due to the fact that they may be able to contribute to both. This causes KMeans to create a "super cluster", or a single cluster which consumes 50\% of the data. This is evident by plotting the distribution of clusters, as seen in Figure 13.

Furthermore, our approach to this problem is inherently flawed. We are attempting to thread errors together based off of their contents, rather than their temporal associations. The purpose behind our original investigation was to determine that, if we forced fewer clusters than would naturally occur, that the errors would coalesce in some way which may provide increased insight. Examining this data, it is very obvious that clustering words based off of their contents will yield similar messages, and we can dismiss this approach as valid and interesting, but not a solution to our problem. In order to re-attack the problem of finding temporal associations, let us examine the abilities of Long-Short Term Memory Networks.

\subsection{Long-Short Term Memory Networks.}
Going back to our review of how Long-Short Term Memories work, we understand that they model long-term dependencies. We hope to leverage this ability by stacking layers composed of LSTM units, which permits any particular neuron to specialize to any given task. Therefore, we hope that each LSTM unit may specialize to some particular type of error, and we are therefore able to leverage this infrastructure to more intelligently group errors, to try and understand which errors may cause some particular error $x$, and which potential errors $x$ may be produced in the future. Finally, our goal is to perform these computations efficiently. If we are able to meet these metrics, then we have successfully proven LSTMs to have some potential which should be further explored.

\begin{figure}
    \centering
    \includegraphics[width=0.4\textwidth]{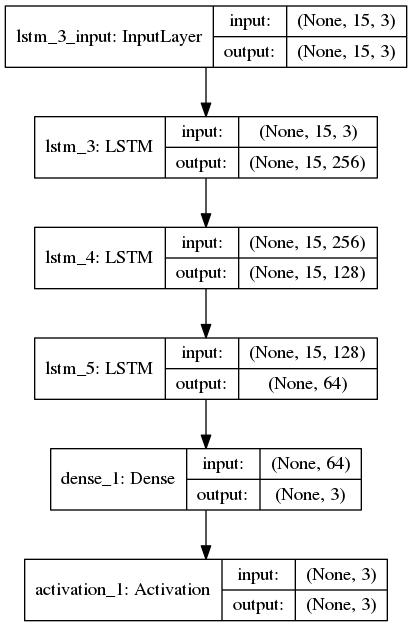}
    \caption{An example model architecture for our deep neural network.}
    \label{fig:my_label}
\end{figure}
Our input to the LSTM is a sequence of $n$ messages, prepared as described in the methodology section of this report. Experimentally, we have determined that an ideal sequence length is fifteen, and that anything larger increases training time without a significant increase in accuracy. However, this has not been rigorously tested.

We train each LSTM for a maximum of 20 iterations, and feed it exactly 128 messages per timestep (batch size). We have implemented Early Stopping, such that the model may finish training if the training loss (a representation of underlying error) remains within 0.00001 of the previous value, and also save the model weights per epoch in case we want to retrieve them later. In Tensorflow, these are implemented with the EarlyStopping and ModelCallback classes respectively. We set aside 20\% of our dataset to validate our findings, and occassionally set the \textit{class\_weight} parameter as described previously.

We learned that the \textit{class\_weight} parameter helps the model learn severely underrepresented classes much better, but takes a large hit in accuracy. We would advise that this parameter is useful if you want to learn a wide range of errors, which is presumably our goal. Therefore, this parameter would be worth inspecting going forward.

\begin{figure}
    \centering
    \includegraphics[width=0.47\textwidth]{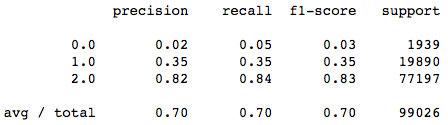}
    \caption{A classification report on our model. This is purely shown to help understand the fact that some hyper parameter tuning is required, but the model shows early promise.}
    \label{fig:my_label}
\end{figure}
In training our network, we had achieved an accuracy of 64\%, and a consistent validation accuracy of 73\%. On average, predictions on our model take about 10 seconds for 100,000 data points. This scales linearly with respect to the data points on hand, therefore meeting our requirements for a classifier faster than $O(n^2)$. However, accuracy for our task is not defined by percentage accuracy, but rather how well we are able to demonstrate a deeper understanding of our dataset, and be able to visualize this understanding. The intuition here is that the ability to predict the next five minutes of syslog messages is relatively useless for us, but the ability to generate abstract clusters, or understand causality is quite useful.

As our goal was to determine whether LSTM cells are able to specialize to any particular type of error or not, we next attempt to understand the activations of these neurons. In other words, we associate which gets the most "excited" about the message with the input message. We then proceed to cluster our messages based off of which neuron was most "excited" by it.

We begin this process by removing the final \textit{Activation} and \textit{Dense} layers, and then recompiling our model. We also determine that it was necessary to manually modify the last final layer and output of the model, which were not changed as a result of removing layers. This process provided a model, which when asked for predictions, returned the activation of each hidden unit in the last layer. This represents how much each unit in the last layer \textit{cared about our syslog messages}, which is a good proxy for unit specialization.

\newpage
\onecolumn
\begin{figure}
    \centering
    \includegraphics[width=\textwidth]{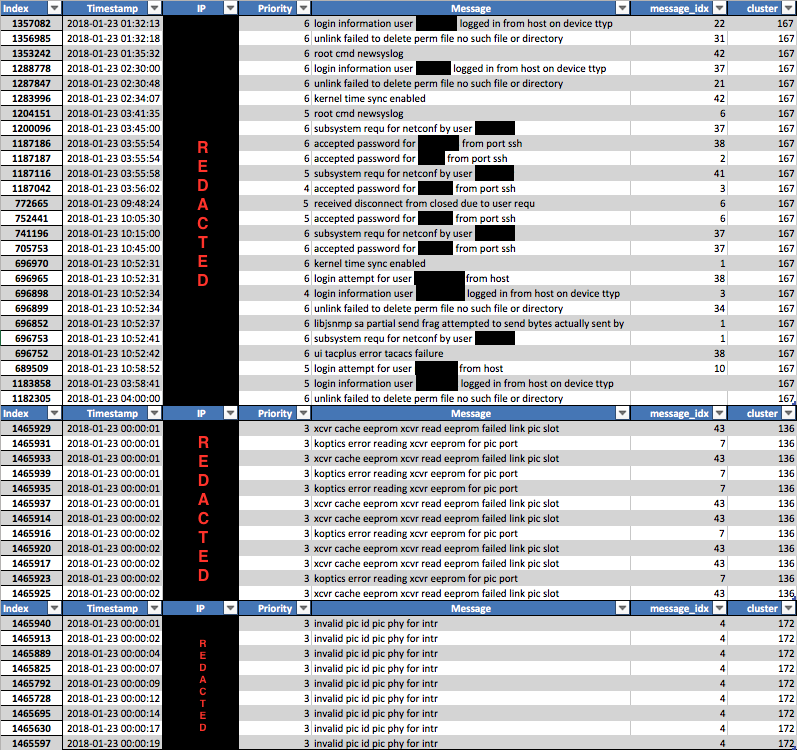}
    \caption{A table visualizing messages being clustered by which hidden unit activated the most.}
    \label{fig:my_label}
\end{figure}
\twocolumn 
Figure 16 shows the results of clustering any particular set of messages based off of each neuron's activation. Most importantly, these vectors \textbf{had no information about the words in the message.} The fact that similar messages are being grouped shows that each \textbf{hidden neuron is understanding the underlying syslog data}. In Cluster 167, we see that messages relating to users are being grouped together. This means that Unit 167 in our LSTM decided that it will be responsible for any messages related to logging in, Secure Shell, etc. On the other hand, Unit 136 understood that some of these messages are related to reading the eeprom, and decided to specialize to this purpose. Once again, these neurons don't know the \textit{content} of each message, just only \textit{when each message occurs.} Occasionally, we see units that will specialize just to one type of message, as evidenced in Unit 172. This shows the future work left to be done. 

Next, we should discuss what this means from a model architecture perspective, such that we may pass these learned intuitions on for others to follow and replicate. We learned that many of our theses about how LSTMs may behave with respect to time series syslog data are correct: \textbf{hidden units do specialize to understand some abstract, higher level error}, and perform better than K-Means and many clustering algorithms, since it is able to group together messages with completely different words. This helps solve one of our first qualms about using LSTMs to understand errors: now we can group errors by \textit{which errors actually occur}, rather than our belief of \textit{which errors we believed to occur.} 

Secondly, we have confirmed that models do \textit{focus} the way we thought they do. Removing hidden (LSTM) layers makes each cluster become more "blurry", these message groupings become more generalized when we have less layers. Similarly, adding additional layers helps the model understand a better representation, and each message grouping becomes more "focused" towards some specific type of error. 

This intuition is also paralleled in the number of hidden units. Beginning with a large number of hidden units, and slowly removing units in each subsequent layer (Fig 14.) is a great method to use to compress information; it discards some useless groupings while maintaining important information. Therefore, it is clear that these networks are able to quickly understand our data, and provide the ability for us to understand their behavior and learn what it is learning.

Next, we will determine if we can understand which previous messages are the most important in predicting the next message. If we are able to accomplish this task, it will permit us to understand what causes some message $x$ to occur, and which messages $x$ may then create. In other words, we may better understand temporal associations in our data.

One issue in the current state of the art is that Long-Short Term Memory networks are good at modeling patterns in long-term dependencies, but not necessarily good at understanding \textit{which} inputs are important in predicting the output. This has lead to concerns in the industry: how will we productionize prediction mechanisms if we can't understand why we predicted it?

Researchers at Carnegie Mellon and Microsoft \cite{Yang2016} have solved this task by implementing something called an \textit{attention mechanism}, which helps the model learn which inputs are able to lead to which outputs by appending random integers to the input, and understand how these integers change as the network learns. As soon as the network is finished learning (putting weights on) the data, we are able to examine which syslog message had the highest weight, and hence understand \textbf{which syslog messages were the most relevant}.

\begin{figure}
    \centering
    \includegraphics[height=\textheight]{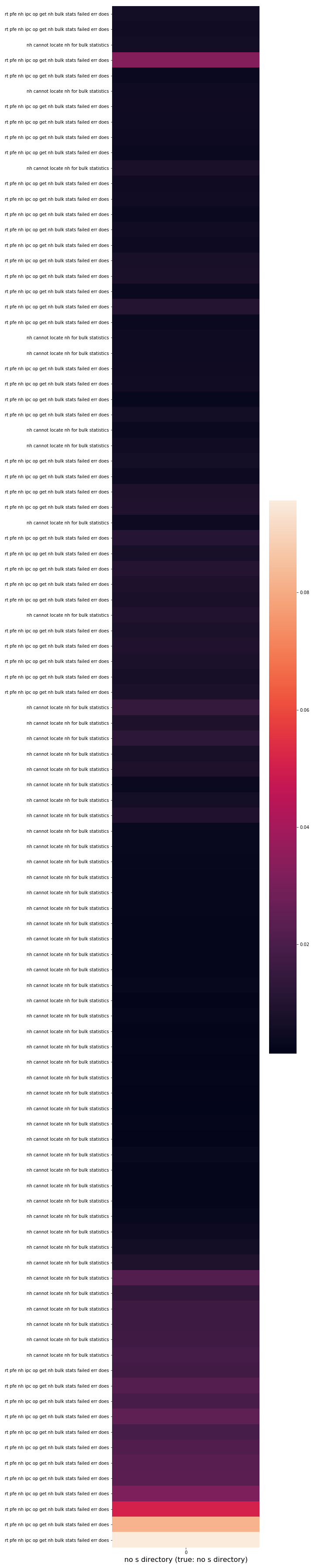}
    \caption{The weight of each previous message, visualized as a heat map. The color bar on the right demonstrates that the more white a message is, the more weight it has on the output. Each row represents a moment in time, so these are how messages are changed over time.}
    \label{fig:my_label}
\end{figure}

Figure 17 on the subsequent page provides an interesting visualization of these weights on the 100 messages directly preceding the output. First, it is evident that most of these messages are not important in determining the output, since most of these messages are colored black. We may use heatmaps like this to determine the optimal sequence length for our model, where we are optimizing for the shortest sequence length that when increased, provides marginal increases in accuracy.

This section of the work is ripe for further exploration. In particular, many other researchers have leveraged attention mechanisms to increase the accuracy of their results, which we lacked the time to implement. Additionally, it remains to be seen whether we can understand which groups of errors may happen next, and at what level of probability. 

In either case, we may consider our work with Long-Short Term Memory Networks to be successful. We have successfully demonstrated that they are able to learn intelligent underlying representations of our data, are able to create meaningful groups of syslog messages from a higher level, and also are able to understand causality between messages. Most importantly, we are able to do all of this under 20 seconds!
\section{Conclusion \& Future Work}
The objective of our work was to attempt to understand the potential that K-Means and Long-Short Term Memory Networks may have on classifying and clustering system log messages. We have shown that K-Means may be used for message classification, but it holds little potential as far as creating a more high-level understanding of errors go. As we have shown in the last section, Long-Short Term Memory Networks hold great promise in this area, and are able to meet our criteria for successful results. They are able to understand the meaning behind each message without knowing what is inside of it, and as a result, they are able to create more powerful and more intelligent clusters than any other algorithm we have previously explored.

The impact of this work could be substantial; it has the potential to enable a significantly more intelligent log analysis frameworks than previously before. We may have the ability to predict whether large, devastating impacts to the network may happen, and provide engineers with a warning beforehand. We may be able to identify a greater variety of errors in our network than previously before, which will permit us to diagnose errors we had never seen before. This will prove paramount in increasing network quality.

However, there is a lot of work to be done: these Long-Short Term Memory Networks must be generalized across a variety of routers. It remains to be seen whether these hyperparameters (number of neurons, learning rate, etc) may be tuned enough such that this process will be ready for production. Additionally, attention should also be implemented as a method of increasing model accuracy, as well as informing end-users of potential errors that have a high probability of occurring.
\section{What I Learned \& Acknowledgements}
This has been one of the most challenging, yet fulfilling internships I have completed to date. I felt quite trapped at the very beginning, as I was unsure whether or not I would be delivering meaningful results, but I'm happy we were able to move in directions which may unlock some potential for the company. Additionally, I'm glad I finally got a chance to implement the theory I had always learned in class; I have been studying the theory behind these notions for a while now, and had been simply itching to get some code down for this.

On a more soft-skills level, I learned a lot about the corporate world during this internship. Comcast is a lot larger than my previous employers, and I actually enjoyed it much more than anywhere I had worked previously.

Finally, I'd like to thank Vibhor Nigam, Patrick Carretas, Dimosthenis Anagnostopoulos, Kim Hall, Phillip Heller, and Tony Tauber for the wonderful company they have provided me during my time here. This experience wouldn't have been the same without them. An additional thank you goes to Vibhor, Dimos, and Tony for providing unending support, and answering my many questions at odd hours.

\nocite{*}
\bibliographystyle{ieeetr}
\bibliography{main}

\begin{thebibliography}{1}

\bibitem{Yang2016}
Z.~Yang, D.~Yang, C.~Dyer, X.~He, A.~Smola, and E.~Hovy, ``{Hierarchical
  Attention Networks for Document Classification},'' in {\em Proceedings of the
  2016 Conference of the North American Chapter of the Association for
  Computational Linguistics: Human Language Technologies}, pp.~1480--1489,
  2016.

\bibitem{Turner2013}
D.~Turner, K.~Levchenko, S.~Savage, and A.~C. Snoeren, ``{A comparison of
  syslog and IS-IS for network failure analysis},'' in {\em Proceedings of the
  2013 conference on Internet measurement conference - IMC '13}, 2013.

\bibitem{Turner2010}
D.~Turner, K.~Levchenko, A.~C. Snoeren, and S.~Savage, ``{California Fault
  Lines: Understanding the Causes and Impact of Network Failures},'' {\em
  Proceedings of the ACM SIGCOMM 2010 Conference}, 2010.

\bibitem{Thompson1968}
K.~Thompson, ``{Regular expression search algorithm},'' {\em Communications of
  the ACM}, vol.~11, no.~6, pp.~419--422, 1968.

\bibitem{Neath2012}
A.~A. Neath and J.~E. Cavanaugh, ``{The Bayesian information criterion:
  Background, derivation, and applications},'' {\em Wiley Interdisciplinary
  Reviews: Computational Statistics}, vol.~4, no.~2, pp.~199--203, 2012.

\bibitem{Vukobratovic2014}
D.~Vukobratovic, C.~Khirallah, V.~Stankovic, and J.~S. Thompson, ``{Random
  network coding for multimedia delivery services in LTE/LTE-Advanced},'' {\em
  IEEE Transactions on Multimedia}, 2014.

\bibitem{Sterbenz2014}
J.~P.~G. Sterbenz, D.~Hutchison, E.~K. {\c{C}}etinkaya, A.~Jabbar, J.~P.
  Rohrer, M.~Sch{\"{o}}ller, and P.~Smith, ``{Redundancy, diversity, and
  connectivity to achieve multilevel network resilience, survivability, and
  disruption tolerance invited paper },'' {\em Telecommunication Systems},
  vol.~56, pp.~17--31, may 2014.

\end{thebibliography}




\end{document}